\documentclass{ws-procs9x6}

\def \bea{\begin{eqnarray}}
\def \eea{\end{eqnarray}}

\begin{document}

\begin{flushright}
YITP-SB-10-29
\end{flushright}

\title{FIXED ANGLE SCATTERING\\
AND THE TRANSVERSE STRUCTURE OF HADRONS}

\author{GEORGE STERMAN}

\address{C.N. Yang Institute for Theoretical Physics, Stony Brook University\\
Stony Brook, New York, 11794-3840, USA\\
$^*$E-mail: george.sterman@stonybrook.edu}

\begin{abstract}
The perturbative treatment of high-energy fixed-angle hadron-hadron exclusive scattering
is reviewed and related to the transverse structure of the proton
and other hadrons.  
\end{abstract}


\bodymatter

\section{Introduction}

Hadron-hadron exclusive cross sections offer a 
sensitivity to transverse partonic substructure that is complementary to most 
photon-induced processes \cite{Ji:1998pc}.
As we shall see, at high enough energies and 
momentum transfers, these cross sections probe hadronic 
Fock states with the minimal numbers of partons, the ``valence" states.  
This talk reviews classic results  on hadron-hadron exclusive 
reactions from this point of view \cite{Matveev:1973ra,Brodsky:1974vy,Landshoff:1974ew,Efremov:1979qk,Lepage:1980fj,Mueller:1981sg}
and recalls as well some related subsequent work \cite{Botts:1989kf,Sotiropoulos:1994ub,Sotiropoulos:1995xy}.

A full treatment of the transition to asymptotic behavior
requires both perturbative resummation and 
nonperturbative input on the transverse
structure of the hadronic valence states.  
It is not really known to what extent 
experiments have yet reached a truly asymptotic 
regime, and pictures involving 
higher Fock states may offer alternative 
descriptions.    A challenge for the future is 
to find a unified, perhaps dual, description of exclusive reactions.
This short presentation, however, will concentrate on the classic
picture based on valence states.   

The next section
reviews the origin of the basic parton-model ``quark counting" 
\cite{Matveev:1973ra,Brodsky:1974vy} predictions for
elastic scattering, grounding them in a simple geometric picture,
in which transverse structure plays only a passive role.
Section \ref{split} introduces the ``Landshoff mechanism", \cite{Landshoff:1974ew}
which reintroduces a dynamic role for transverse
structure, and Sec.\ \ref{res} shows how an analysis
of radiative corrections links the two.   In Sec.\ \ref{hhbar},
some intriguing data on wide-angle particle-antiparticle
scattering is briefly discussed.

\section{Valence states, geometry and quark counting}
\label{geo}

\begin{figure}
\begin{center}
\psfig{file=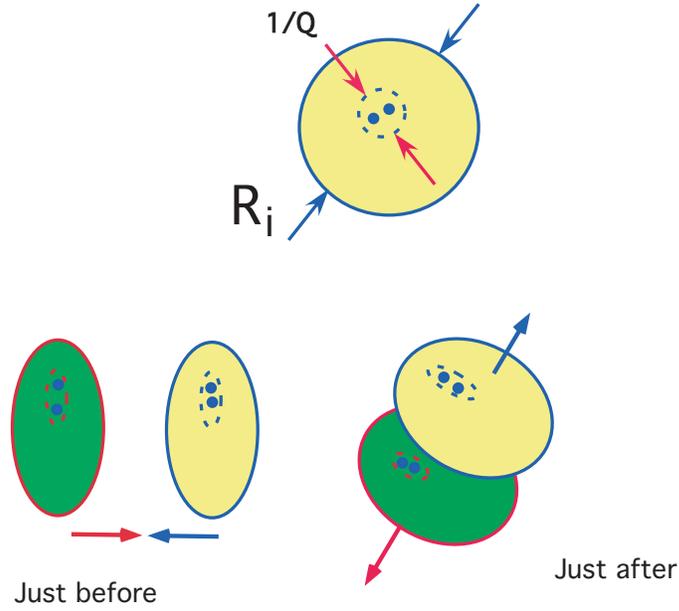,width=3.5in}
\end{center}
\caption{Valence states before and after an elastic scattering.}
\label{fig: counting}
\end{figure}

\begin{figure}
\begin{center}
\psfig{file=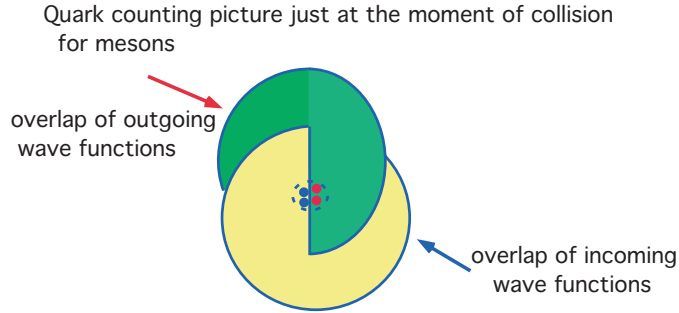,width=3.5in}
\end{center}
\caption{Incoming and outgoing wave functions at the moment of collision.}
\label{fig: single}
\end{figure}

We begin with the parton model applied to high-energy elastic scattering.
In Refs.\ \refcite{Matveev:1973ra} and \refcite{Brodsky:1974vy}, 
elastic scattering is pictured as occurring via time-dilated Fock
states with fixed numbers of partons.
Suppose all  the $n_H$ partons in the valence state for hadron $H$ 
have comparable  momentum fractions $x_i$. 
A large coherent momentum transfer $t\sim -Q^2$ is necessary to redirect all these partons into
another direction.   Such a momentum transfer 
requires all of the $n_H$ (anti-)quarks to be
in a region of area $1/Q^2$ in the Lorentz-contracted wave functions
of the colliding hadrons in the center-of-mass frame.
This has to be the case for each colliding hadron,
and also for the two hadrons that emerge from the
scattering.
The essential observation is that if the
distribution of partons in the transverse
direction is random, the likelihood for such a configuration
 is estimated by $\left( \frac{1}{Q^2}\ \times\ \frac{1}{\pi R_H^2} \right)^{n_H-1}$
for each hadron.   This geometric picture is
illustrated in Figs.\ \ref{fig: counting} and \ref{fig: single}.

If the geometric picture determines the overall energy and momentum
dependence, then otherwise the amplitude is a function only of the scattering 
angle.   At fixed $s/t$ (that is, fixed c.m.\ scattering angle) we find,
\bea
\frac{d\sigma}{dt} = \frac{f(s/t)}{s^2}\ \left ( \frac{m^2}{s}\right)^{\sum_{i=1}^4\, (n_i\, -\, 1)}\, .
\eea
This is the basic parton model result, the `quark counting rule'.   It clearly picks out the valence state:
the cross sections associated with larger numbers of partons are power suppressed,
at least as long as we assume that they do not have vanishing momentum fractions.

As the parton model took on a new life in the late
1970's as a limit of quantum chromodynamics, 
it was discovered in groundbreaking papers
 that the corresponding  elastic amplitude 
 could be written as
\cite{Efremov:1979qk,Lepage:1980fj}
\bea
{\cal M}(s,t;h_i)
&=&
{\int \prod_{i=1}^4 \ [dx_i]}\
\phi(x_{m,i},\lambda_{m,i},h_i;\mu)
\nonumber\\
&\ & \times
M_H \left (\frac{x_{n,i}x_{m,j}p_i\cdot p_j}{\mu^2};\lambda_{n,i},h_i\right)\, ,
\label{eq:xamp}
\eea
in terms of a calculable hard-scattering amplitude
$M_H$ in convolution with
factorized and evolved valence (light-cone) wave functions,
 $\phi(x_{m,i},\lambda_{m,i},h_i;\mu)$,
   with helicities $h_i$ for hadrons and $\lambda_{n,i}$ for quarks.   
   The mass $\mu$ is the factorization scale.    The convolution
 is in terms of the partonic momentum fractions: for baryons,
\bea
[dx_i] = dx_{1,i}dx_{2,i}dx_{3,i}\, \delta\left( 1 - \sum_{n=1}^3 x_{n,i}\right)\, .
\eea
So far, the transverse dimensions of colliding hadrons 
enter only indirectly, in the geometric justification of power behavior.

In principle, all this is straightforward, but our knowledge of the
wave functions is not complete, and in any case for nucleon-nucleon
scattering there are too many diagrams even at low orders to make
a direct calculation practical, at least up to this time.

\section{Splitting the hard scattering}
\label{split}

\begin{figure}
\begin{center}
\psfig{file=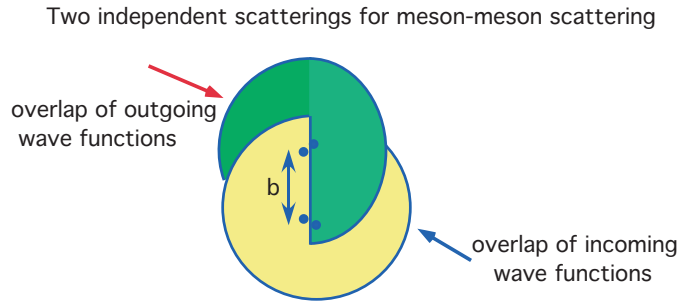,width= 3.5 in}
\end{center}
\caption{Representation of independent scatterings.}
\label{doublescattering}
\end{figure}

The geometric analysis for elastic scattering in the valence
state is more flexible than it might at first seem, and there is an alternative
picture of the rearrangement of parton momenta, shown in Fig.\ \ref{doublescattering}.
This is the geometric interpretation of the process identified first by
Landshoff \cite{Landshoff:1974ew}.

Figure \ref{doublescattering} shows that the single hard scattering
of quark counting can in principle be split into as many
hard scatterings as there are valence partons in each of the incoming
hadrons.   (The figure shows two, as for $\pi\pi$ scattering.)   
Each parton must overlap with an incoming and outgoing wave function.
Relative to the single hard scattering picture, however, we no
longer require all of the partons to be in the same region
of radius $1/Q$.   If there are, for example, two hard scatterings,
we gain a geometric factor of the transverse size of the
hadrons, $R_HQ$ relative to single scattering, with $Q$
the momentum transfer.
An enhancement $1/Q \rightarrow  R_H$ 
in the amplitude gives a factor $1/Q^2 \rightarrow R_H^2$ in cross section.
In this way, the Landshoff analysis provides an enhancement by
a factor $Q^2\sim s$ for fixed angle pion scattering cross sections and $Q^4\sim s^2$
for proton-proton fixed angle elastic scattering.
More specifically, this modified geometric configuration gives\cite{Landshoff:1974ew}
 for pp at fixed angle ({\it i.e.} fixed $s/t$)
\bea
\frac{d\sigma}{dt} = \frac{f(s/t)}{s^2}\ \left ( \frac{1}{s\, \pi R_H^2}\right)^6\, ,
\eea
while for $s\gg -t \gg \Lambda_{\rm QCD}$ it gives
\bea
\frac{d\sigma}{dt} = \frac{F(s)}{t^2}\ \left ( \frac{1}{t\, \pi R_H^2}\right)^6\, .
\eea
Experimentally, the forward scattering proposal works well at
a wide range of center-of-mass energies, but
at fixed angles the data appear to follow the original quark counting rules.
As I'll now argue, this distinction may be associated with the
transverse substructure of hadrons.

\section{The return of (approximate) quark counting
at wide angles}
\label{res}

\begin{figure}
\begin{center}
\psfig{file=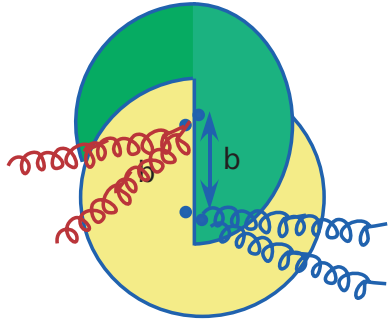,width=2 in}
\end{center}
\caption{Schematic representation of radiation that is absent
in elastic scattering.}
\label{doublerad}
\end{figure}

The scattering of isolated color charges tends to produce radiation in the incoming
and outgoing directions.  
Figure \ref{doublerad} illustrates this effect, superimposed on the
parton model template of Fig.\ \ref{doublescattering} for pion scattering.
For finite impact parameter $b$, each of the two scattering processes along the
vertical line of overlap involves  colored partons.    This should 
lead to the radiation of gluons of wavelength as small as 
order $1/Q$ and as large as $b$.
Final states without such radiation are suppressed by virtual corrections,
unless the impact parameters $b$ is small.   
On the one hand the Landshoff mechanism is enhanced because there are more ways to produce
two hard scatterings than one, but on the other hand it is suppressed because
isolated scatterings of colored particles are only rarely elastic.
The full amplitude is actually the result of a competition between
geometric enhancement and radiative suppression. \cite{Mueller:1981sg,Pire:1982iv}   
The resulting balance, and its energy and momentum transfer dependence,
was analyzed in Ref.\ \refcite{Botts:1989kf}.

\begin{figure}
\begin{center}
\psfig{file=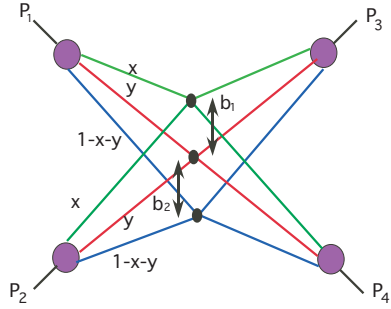,width=2 in}
\end{center}
\caption{The Landshoff mechanism in impact parameter space.}
\label{fig:angular}
\end{figure}

Impact parameter $b$ is conjugate to $Q=\sqrt{-t}$.   As $-t$ increases toward $s$, radiative corrections
force the $b$'s to $1/\sqrt{s}$ and the quark counting geometric picture should be  recovered approximately.
\cite{Lepage:1980fj, Mueller:1981sg,Landshoff:1980pb}
 For color-singlet nucleons, we have as many as three independent scatterings,
 and this sort of analysis leads for baryons to  the expression (see Fig.\ \ref{fig:angular}) \cite{Sotiropoulos:1995xy}
\bea
{\cal M}(s,t)
&=&
\frac{N}{stu} \sum_f \int_0^1 \frac{dx dy}{x^2y^2(1-x-y)^2}\  
\nonumber\\
\nonumber\\ 
&\ & \hspace{5mm} \times
\int db_1db_2\  {\rm Tr}_{\rm color}\left[U(b_iQ)M^1M^2M^3\right]
\nonumber\\
&\ & \hspace{20 mm} \times\ \prod_{m=1,2,3,4}\ \Psi_{H_m}(x,y,b_1,b_2)\, ,
\label{eq:ampfact}
\eea
where $N$ is a normalization factor.
 The Trace $\left[U(b_iQ)M^1M^2M^3\right]$ ties color together and includes
$\epsilon_{abc}$ for colors of three quarks in each external hadron, with possible color exchange in
each hard scattering $M^i(x_ip_j)$.  The $M$s are independent
of the transverse separations.   The  matrix $U$, which
depends on both transverse distances $b_i$, is necessary 
to incorporate coherent logarithms, left over after the
factorization of the wave functions, $\Psi$.

  The wave functions depend on the center-of-mass
  energy, rather than a factorization scale, as would be
  the case for parton distributions.   This is 
  because they represent the creation or absorption of
  the partons of the valence state  at
  equal light cone time but  fixed 
  transverse separation.  In this way, they reflect the amplitude
  for such a state to exist, without additional 
  degrees of freedom, in a particular frame. They 
  are not gauge invariant, but their product in Eq.\ (\ref{eq:ampfact}) is invariant.
  In fact, it is convenient to construct the wave functions in a physical gauge \cite{Collins:1981uk} 
  and to use their gauge dependence to determine their energy dependence.
The pattern is illustrated by the simpler example of pion wave functions, for which
one finds an exponentiated form,\cite{Botts:1989kf}
\bea
\Psi(x,b,Q) &\sim& \Psi_{NP}(x,b)\exp[-s(x,b,Q) -s(1-x,b,Q)] \, ,
\eea
where the two partons of the valence state carry fractional momenta $x$ and $1-x$,
where $Q\equiv \sqrt{-t}$  and $b$ is the distance between the hard scatterings in
Fig.\ \ref{doublerad}.   The functions $s(\xi,b,Q)$ provide double-logarithmic
suppression whenever the product $Qb$ is large, being defined 
in terms of universal anomalous dimensions.   These
are  $A(\alpha_s)=C_F(\alpha_s/\pi)+ \dots $, which organizes all leading
logarithms, and $B(\alpha_s)$, which depends on the details of the factorization
procedure,
\bea
s(x,b,Q) = \int_{C_1/b}^{C_2xQ} \frac{d\mu}{\mu} \left [
4\ln\left(\frac{C_2xQ}{\mu}\right)A(\alpha_s(\mu)) + B(\alpha_s(\mu))
\right]\, .
\label{eq:sdef}
\eea
For fixed coupling and small $b$, one finds that the wave functions have double-logarithmic dependence on the impact parameters,
\bea
\Psi(x,b,Q) &\rightarrow& \phi_{asy}(x_j)\exp[- {\rm const}\; \ln^2(1/Qb)]\, .
\eea
Assembling the pieces for baryon scattering,
this gives the following asymptotic amplitude, with an example of ``Sudakov resummation"
when $b\gg 1/Q$, which drives the wave functions to their `asymptotic' behaviors\cite{Lepage:1980fj,Botts:1989kf}
\bea
{\cal M}(s,t)
&=&
\frac{N}{stu} \sum_f \int_0^1 \frac{dx_1 dx_2}{x_1^2x_2^2(1-x-y)^2}\  \prod_{m=1,2,3,4}\ \phi_{m,asy}(x,y)
\nonumber\\
\nonumber\\ 
&\ & \hspace{5mm} \times
\int db_1db_2\  {\rm Tr}_{\rm color}\left[U(b_iQ)M^1M^2M^3\right]
\nonumber\\
\ \nonumber\\
&\ & \hspace{20 mm} \times\ e^{-S_1(b_iQ)-S_2(b_iQ)-S_3(b_iQ)}\, .
\label{eq:Mst}
\eea
 At large $Q$ for each scattering, radiation suppression also forces the 
hard scatterings back together.   The $S_i$ are sums of exponential functions like those in (\ref{eq:sdef}),
one for each scattering.
 At moderate $(xQ)^2,\, (yQ)^2$, the amplitude is dominated by the ``boundary conditions,"
$\Psi_{NP}(x,y,b_i)$ rather than asymptotic behavior.   One of the attractive features of
this expression is that the original eight integrals over momentum fractions $x_{m,i}$ 
in Eq.\ (\ref{eq:xamp}) are reduced to only two.

An important feature of Eq.\ (\ref{eq:Mst}) is that the scale of Sudakov suppression is set by
the momentum transfer.   Varying $t$ at fixed $s$ thus in principle modifies the role of
radiative corrections, and as $|t|$ decreases, we anticipate that the cross section 
makes a transition from $s^{-10}$ to $t^{-8}$ behaviour, as seen in the data \cite{Nagy:1978iw}.
A study of the implications of Eq.\ (\ref{eq:Mst}) for the transverse size of the proton
valence state was made in Ref.\ \refcite{Sotiropoulos:1995xy}, where it was concluded that the data favor
a quite small transverse extent.   Fixed-angle scattering on nuclei offers another 
possible test of the transverse structure of hadrons,
via nuclear transparency for small-size color-singlet states \cite{Bertsch:1981py,Jain:1995dd,Frankfurt:1993it,Carroll:1988rp}.

\section{Exchanging quarks}
\label{hhbar}

The formalism we've described so far is valid asymptotically, but
it is not so clear how high  the energy and momentum transfer have to
be for corrections to these pictures to be negligible.   For the multiple
scattering picture, in particular, the invariant mass for the ``hard scattering"
described by the amplitude $M_i$ in Eq.\ (\ref{eq:Mst})
is $x_i^2s$, typically an order of magnitude smaller than 
the hadron-level invariants \cite{Isgur:1988iw}.   At the very least, it is clear that the
wave functions must vanish sufficiently fast when any of the $x_i\rightarrow 0$
in the valence state.   Considerations such as these suggest that
at accessible energies, alternative descriptions of exclusive 
scattering, not necessarily limited to the valence state, should
be examined.    For the closely related case of form factors, descriptions based on QCD sum
rules have shown success\cite{Nesterenko:1982gc}.

With all this in mind, and because of the complexity of the calculations involved, 
further experimental comparisons of elastic amplitudes between different hadrons may lead to
valuable insights.
For example, a conceptual contrast was made between gluon and quark exchange
processes early on, and has remained of interest \cite{Gunion:1973ex,Ramsey:1994ay}.   
Quark exchange should be highly sensitive to flavor content.

The perturbative factorization formalisms described above can be thought of
as involving the exchange of quark degrees of freedom in 
addition to gluons, and generally these processes contribute to the amplitude in a complicated way.
A case of interest, however, is the comparison
of $p \bar p$ to $pp$ scattering.
For $pp$, there are $2^3$ ways of connecting incoming and outgoing 
quarks compared to only one for $p\bar p$.   If amplitudes are coherent sums
of these processes, all with a similar weight, then we might expect a ratio of $1/64$ for the elastic
scattering cross sections of protons on antiprotons to that for protons on protons.  

The relevance of this elegant observation was tested in experiments
at  Brookhaven National Laboratory's AGS \cite{Blazey:1985pq}, 
and ratios of particle-antiparticle to particle-particle seemed roughly consistent with this counting,
$
R_N = \frac{\frac{d\sigma_{N\bar N}}{dt}}{\frac{d\sigma_{N N}}{dt}}\, \Bigg |_{90\ \rm deg}
\sim
\frac{1}{40}
$.  
Any successful picture of these exclusive hadronic processes must explain this
simple counting result.

Sotiropoulos\cite{Sotiropoulos:1995xy} studied this issue in the Landshoff multiple scattering
picture.    At first sight, the situation seems promising:  at each
hard scattering in Fig.\ \ref{fig:angular}, quarks may be exchanged (or not)
between the participating protons of $pp$ scattering in all possible
ways, but for $p\bar p$ scattering, the quarks and antiquarks have
only one way to get from the initial to the final state.
There is a caveat in any pQCD picture, however:   we have to reform an antisymmetric 
color combination of quarks when they are exchanged, a process that requires them to
exchange gluons.
Sotiropoulos found that the pertubative picture sketched above works qualitatively 
only with a ``color randomization"
hypothesis in which the factor $\left[U(bQ)\prod_iM^i\right]$ 
is independent of the flavor flow.   He found a ratio of cross sections of $R_p\sim 1/30$ 
at ninety degrees with color randomization, but only $\sim 1/3$ without it,
with a qualitatively successful angular dependence\cite{Farrar:1974jz,Sotiropoulos:1995xy}.
Randomization is a plausible effect at the moderate BNL energies, $\sqrt{s}\sim 3.5$ GeV$^2$,
and is easy to picture in the context of quark exchange.
This leaves us with a description that is a mixture of
perturbative and nonperturbative effects.
 On the other hand, 
 the Landshoff/Sudakov model with or without randomization
 gives explicit predictions for angular and helicity dependence.
 Clearly, a decrease of $R_{pp}$ with energy would be a compelling signal for
an emerging role for color.

\section{Conclusions}

It has been some time since hadron-hadron elastic scattering 
has held center stage.   It remains, however, an important probe
of hadron structure, especially for information on pion and
nucleon valence states.   New facilities may push the frontier
in $s$ and $t$ coverage, and clarify essential questions
of the role of perturbative and nonperturbative scattering
mechanisms.   Insights flowing from the
development of duality-based pictures of hadronic structure \cite{Brodsky:2006uqa,Grigoryan:2008cc}
and the increase in our capability to compute multi-parton
scattering amplitudes \cite{Bern:2007dw} should provide new tools for 
these fundamental processes.

\section{Acknowledgments} 

This work was supported in part by the National Science Foundation, grant
PHY-0653342.   I thank Mark Strikman for discussions, and the organizers
of ``Exclusive Reactions at High Momentum Transfer",
especially 
Anatoly Radyushkin, Paul Stoler and Christian Weiss, for the invitation to speak
at the workshop


\begin{thebibliography}{9}

\bibitem{Ji:1998pc}
  X.~D.~Ji,
  J.\ Phys.\ G {\bf 24}, 1181 (1998)
  [arXiv:hep-ph/9807358].
  A.~V.~Belitsky and A.~V.~Radyushkin,
  Phys.\ Rept.\  {\bf 418}, 1 (2005)
  [arXiv:hep-ph/0504030].

\bibitem{Matveev:1973ra}
  V.~A.~Matveev, R.~M.~Muradian and A.~N.~Tavkhelidze,
  {\it Lett.\ Nuovo Cim.}\  {\bf 7}, 719 (1973).

\bibitem{Brodsky:1974vy}
  S.~J.~Brodsky and G.~R.~Farrar,
  {\it Phys.\ Rev.}\  D {\bf 11}, 1309 (1975).
  
\bibitem{Landshoff:1974ew}
  P.~V.~Landshoff,
  {\it Phys.\ Rev.\  D} {\bf 10}, 1024 (1974).
  
\bibitem{Efremov:1979qk}
  A.~V.~Efremov and A.~V.~Radyushkin,
 {\it  Phys.\ Lett.}\  B {\bf 94}, 245 (1980).
 
\bibitem{Lepage:1980fj}
  G.~P.~Lepage and S.~J.~Brodsky,
  {\it Phys.\ Rev.}\  D {\bf 22}, 2157 (1980).

\bibitem{Mueller:1981sg}
  A.~H.~Mueller,
  {\it Phys.\ Rept.}\  {\bf 73}, 237 (1981).
  
\bibitem{Botts:1989kf}
  J.~Botts and G.~F.~Sterman,
  Nucl.\ Phys.\  B {\bf 325}, 62 (1989).
  
\bibitem{Sotiropoulos:1994ub}
  M.~G.~Sotiropoulos and G.~F.~Sterman,
 {\it  Nucl.\ Phys.}\  B {\bf 425}, 489 (1994)
  [arXiv:hep-ph/9401237].
  
\bibitem{Sotiropoulos:1995xy}
  M.~G.~Sotiropoulos,
 {\it  Phys.\ Rev.}\  D {\bf 54}, 808 (1996)
  [arXiv:hep-ph/9512397].
    
\bibitem{Pire:1982iv}
  B.~Pire and J.~P.~Ralston,
 {\it Phys.\ Lett.}\  B {\bf 117}, 233 (1982).
    
\bibitem{Landshoff:1980pb}
  P.~V.~Landshoff and D.~J.~Pritchard,
  {\it Z.\ Phys.}\  C {\bf 6}, 69 (1980).
  
\bibitem{Collins:1981uk}
  J.~C.~Collins and D.~E.~Soper,
  Nucl.\ Phys.\  B {\bf 193}, 381 (1981)
  [Erratum-ibid.\  B {\bf 213}, 545 (1983)].
  
\bibitem{Nagy:1978iw}
  E.~Nagy {\it et al.},
 {\it  Nucl.\ Phys.}\  B {\bf 150}, 221 (1979).
  W.~Faissler {\it et al.},
{\it  Phys.\ Rev.}\  D {\bf 23}, 33 (1981).

\bibitem{Bertsch:1981py}
  G.~Bertsch, S.~J.~Brodsky, A.~S.~Goldhaber and J.~F.~Gunion,
  {\it Phys.\ Rev.\ Lett.}\  {\bf 47}, 297 (1981);
  S.~J.~Brodsky and A.~H.~Mueller,
  {\it Phys.\ Lett.}\  B {\bf 206}, 685 (1988).

\bibitem{Jain:1995dd}
  P.~Jain, B.~Pire and J.~P.~Ralston,
  {\it Phys.\ Rept.}\  {\bf 271}, 67 (1996)
  [arXiv:hep-ph/9511333].
  
\bibitem{Frankfurt:1993it}
  L.~Frankfurt, G.~A.~Miller and M.~Strikman,
  {\it Phys.\ Lett.}\  B {\bf 304}, 1 (1993)
  [arXiv:hep-ph/9305228];

\bibitem{Carroll:1988rp}
  A.~S.~Carroll {\it et al.},
 {\it  Phys.\ Rev.\ Lett.}\  {\bf 61}, 1698 (1988);
  E.~M.~Aitala {\it et al.}  [E791 Collaboration],
  {\it Phys.\ Rev.\ Lett.}\  {\bf 86} (2001) 4773
  [arXiv:hep-ex/0010044].
  
\bibitem{Isgur:1988iw}
  N.~Isgur and C.~H.~Llewellyn Smith,
  {\it Nucl.\ Phys.}\  B {\bf 317}, 526 (1989).
  
\bibitem{Nesterenko:1982gc}
  V.~A.~Nesterenko and A.~V.~Radyushkin,
{\it   Phys.\ Lett.}\  B {\bf 115}, 410 (1982).
  
\bibitem{Gunion:1973ex}
  J.~F.~Gunion, S.~J.~Brodsky and R.~Blankenbecler,
  {\it Phys.\ Rev.}\  D {\bf 8}, 287 (1973).
  
\bibitem{Ramsey:1994ay}
  G.~P.~Ramsey and D.~W.~Sivers,
 {\it  Phys.\ Rev.}\  D {\bf 45}, 79 (1992).
 {\it Phys.\ Rev.}\  D {\bf 52}, 116 (1995).
 
\bibitem{Blazey:1985pq}
  G.~C.~Blazey {\it et al.},
 {\it  Phys.\ Rev.\ Lett.}\  {\bf 55}, 1820 (1985).
  B.~R.~Baller {\it et al.},
 {\it  Phys.\ Rev.\ Lett.}\  {\bf 60}, 1118 (1988).

  C.~G.~White {\it et al.},
 {\it  Phys.\ Rev.}\  D {\bf 49}, 58 (1994).

\bibitem{Farrar:1974jz}
  G.~R.~Farrar and C.~C.~Wu,
 {\it  Nucl.\ Phys.}\  B {\bf 85}, 50 (1975).
 
\bibitem{Brodsky:2006uqa}
  S.~J.~Brodsky and G.~F.~de Teramond,
  Phys.\ Rev.\ Lett.\  {\bf 96}, 201601 (2006)
  [arXiv:hep-ph/0602252].
 
\bibitem{Grigoryan:2008cc}
  H.~R.~Grigoryan and A.~V.~Radyushkin,
  Phys.\ Rev.\  D {\bf 78}, 115008 (2008)
  [arXiv:0808.1243 [hep-ph]].
  
\bibitem{Bern:2007dw}
  Z.~Bern, L.~J.~Dixon and D.~A.~Kosower,
  Annals Phys.\  {\bf 322}, 1587 (2007)
  [arXiv:0704.2798 [hep-ph]].


\end{thebibliography}
\end{document}